\documentclass[aps, prd, amsmath, floats, floatfix, twocolumn, nofootinbib, superscriptaddress]{revtex4-1}
\usepackage{graphicx}
\usepackage{color}
\usepackage{latexsym}
\usepackage[colorlinks]{hyperref}

\newcommand{\be}{\begin{eqnarray}}
\newcommand{\ee}{\end{eqnarray}}

\begin{document}

\title{Testing conformal gravity with the supermassive black hole in 1H0707--495}

\author{Menglei~Zhou}
\affiliation{Center for Field Theory and Particle Physics and Department of Physics, Fudan University, 200438 Shanghai, China}

\author{Zheng~Cao}
\affiliation{Center for Field Theory and Particle Physics and Department of Physics, Fudan University, 200438 Shanghai, China}

\author{Askar~Abdikamalov}
\affiliation{Center for Field Theory and Particle Physics and Department of Physics, Fudan University, 200438 Shanghai, China}

\author{Dimitry~Ayzenberg}
\affiliation{Center for Field Theory and Particle Physics and Department of Physics, Fudan University, 200438 Shanghai, China}

\author{Cosimo~Bambi}
\email[Corresponding author: ]{bambi@fudan.edu.cn}
\affiliation{Center for Field Theory and Particle Physics and Department of Physics, Fudan University, 200438 Shanghai, China}
\affiliation{Theoretical Astrophysics, Eberhard-Karls Universit\"at T\"ubingen, 72076 T\"ubingen, Germany}

\author{Leonardo~Modesto}
\affiliation{Department of Physics, Southern University of Science and Technology (SUSTech),\\ 
1088 Xueyuan Road, Shenzhen 518055, China}

\author{Sourabh~Nampalliwar}
\affiliation{Theoretical Astrophysics, Eberhard-Karls Universit\"at T\"ubingen, 72076 T\"ubingen, Germany}

\begin{abstract}
Recently, two of us have found a family of singularity-free rotating black hole solutions in Einstein's conformal gravity. These spacetimes are characterized by three parameters: the black hole mass $M$, the black hole spin angular momentum $J$, and a parameter $L$ that is not specified by the theory but can be expected to be proportional to the black hole mass $M$. The Kerr black hole solution of Einstein's gravity is recovered for $L = 0$. In a previous paper, we showed that X-ray data of astrophysical black holes require $L/M < 1.2$. In the present paper, we report the results of a more sophisticated analysis. We apply the X-ray reflection model {\sc relxill\_nk} to \textsl{NuSTAR} and \textsl{Swift} data of the supermassive black hole in 1H0707--495. We find the constraint $L/M < 0.45$ (90\% confidence level).
\end{abstract}

\maketitle


\section{Introduction}

Einstein's gravity was proposed at the end of 1915 and, as of now, has successfully passed a large number of observational tests~\cite{will}. However, while the theory has been tested in weak gravitational fields, the strong gravity regime is largely unexplored. Additionally, there are many alternative theories of gravity that have the same predictions as Einstein's gravity in weak gravitational fields and present deviations only when gravity becomes strong. The ideal laboratory for testing strong gravity is the spacetime around astrophysical black holes~\cite{book}.

In 4-dimensional Einstein's gravity, the only stationary, axisymmetric, and asymptotically flat uncharged black hole solution, which is regular on and outside the event horizon, is described by the Kerr metric~\cite{h1,h2,h3}. The spacetime geometry around astrophysical black holes formed from gravitational collapse is thought to be well approximated by the Kerr solution~\cite{book}. Initial deviations from the Kerr metric are quickly radiated away by the emission of gravitational waves~\cite{price}. Deviations from the Kerr background due to surrounding matter, like an accretion disk or nearby stars, is normally negligible~\cite{k2a,k2b}. For macroscopic black holes, the electric charge can be completely ignored~\cite{bdp,book}. In the end, macroscopic deviations from the Kerr metric can only be possible in the presence of new physics.

Black hole solutions in theories beyond Einstein's gravity have been extensively investigated in the past few decades. They may have different theoretical (e.g. uniqueness of solutions, thermodynamics stability, entropy, topology of the horizon, etc.) and observational properties with respect to the Kerr black holes of Einstein's gravity, and well-known results valid in the standard theory may not hold. See, for instance, \cite{ppp0,ppp1,pp2,ppp2,Berti:2015itd,rmp} and reference therein.

The Kerr black hole hypothesis can be tested by studying either the properties of the electromagnetic radiation emitted by gas or stars orbiting a black hole~\cite{rmp,em} or the gravitational waves emitted by a system with a black hole~\cite{gw1,gw2}. Each method has its advantages and disadvantages, and the two techniques are complementary as they test different physics. Strictly speaking, electromagnetic radiation approaches can test the motion of massive and massless particles in the strong gravity region around a black hole. If we assume geodesic motion, we can test the Kerr metric. The gravitational wave signal is instead determined by the evolution of perturbations over a background metric and are thus sensitive to the field equations of the gravity theory.

There are two natural strategies to test the Kerr nature of astrophysical black holes using electromagnetic radiation, and they are generally referred to as, respectively, top-down and bottom-up approaches~\cite{rmp}. In the former case, we consider a non-Kerr black hole solution of a particular alternative theory of gravity and we use observations to check if astrophysical data prefer the Kerr metric of Einstein's gravity or the non-Kerr metric of the alternative theory of gravity under investigation. With the bottom-up approach, we do not consider a theoretically motivated metric and instead employ a phenomenological test-metric with a number of ``deformation parameters'' that are used to capture possible non-Kerr features. Observational data are then used to measure the values of these deformation parameters and see if the results are consistent with what is expected in the case of the Kerr spacetime.

Many electromagnetic tests of the Kerr metric in the literature rely on the bottom-up approach. The main reason is that rotating black hole solutions in alternative theories of gravity are often unknown, while tests of the Kerr metric are not possible, or more challenging, with non-rotating or slow-rotating solutions~\cite{slow}. This is related to the difficulties in solving the corresponding field equations for rotating solutions and it is evident even in the case of Einstein's gravity. Indeed, the non-rotating Schwarzschild black hole solution of Einstein's gravity was found in 1916, shortly after the theory was proposed. The rotating Kerr black hole solution was found by Roy Kerr in~1963, more than forty years after the discovery of the non-rotating solution.

In the present paper, we want to test the black hole solution of conformal gravity\footnote{While there are already several studies on the astrophysical and cosmological constraints on Weyl's conformal gravity, with controversial claims of consistency and inconsistency of the model with observations~\cite{test1,test2,test3,test4}, the scenario of Einstein's conformal gravity of Ref.~\cite{p1} is currently unexplored with the exception of our tests on the Kerr metric.} found in Ref.~\cite{p1} (see Refs.~\cite{inter1,inter2,inter3,inter4} for the physical interpretation of these spacetimes). This is an exact rotating black hole solution of a large family of conformally invariant theories of gravity. In Boyer-Lindquist coordinates, the line elements reads~\cite{p1}\footnote{Throughout the paper, we employ units in which $G_{\rm N} = c = 1$ and a metric with signature $(-+++)$.}
\be\label{eq-ds}
ds^2 = \left( 1 + \frac{L^2}{\Sigma} \right)^2 ds^2_{\rm Kerr} 
\ee
where $ds^2_{\rm Kerr}$ is the line element of the Kerr metric
\be
ds^2_{\rm Kerr} &=&  - \left( 1 - \frac{2 M r}{\Sigma} \right) \, dt^2
- \frac{4 M a r \sin^2\theta}{\Sigma} \, dt \, d\phi
\nonumber\\ &&
+ \left(r^2 + a^2 
+ \frac{2 M a^2 r \sin^2\theta}{\Sigma}\right) \, \sin^2\theta \, d\phi^2
\nonumber\\ &&
+ \frac{\Sigma}{\Delta} \, dr^2 + \Sigma \, d\theta^2  \, ,
\ee
$\Sigma = r^2 + a^2 \cos^2\theta$, $M$ is the black hole mass, $a = J/M$ is the specific spin, $J$ is the black hole spin angular momentum, and $L$ is a new parameter of the model. $L$ can naturally be either of the order of the Planck length, $L_{\rm Pl} \sim 10^{-33}$~cm, or of the order of the black hole mass, $M$, as these are the only two scales in the system. In what follows, we consider the case in which $L \propto M$, since it is unlikely that the scenario with $L \propto L_{\rm Pl}$ can be tested with astrophysical observations.

In Ref.~\cite{p2}, we calculated the profile of the iron K$\alpha$ line that can be expected in the reflection spectrum of a putative accretion disk in a black hole spacetime described by the line element in Eq.~(\ref{eq-ds}). For high spins, as the parameter $L$ increases, the radius of the innermost stable circular orbit (ISCO) increases as well, and this has clear observational implications in the X-ray spectrum of black holes. In particular, for sufficiently large values of $L/M$, the reflection spectrum of the accretion disk cannot have very broad iron lines. However, very broad iron lines are observed in the X-ray spectrum of several black holes. We simulated 100~ks observations with \textsl{NuSTAR}\footnote{http://www.nustar.caltech.edu} and we obtained the constraint $L/M < 1.2$, arguing that larger values of $L/M$ would have been observed as anomalous features in the available disk's reflection spectra of astrophysical black holes.

In the present work, we want to improve the study in Ref.~\cite{p2} and attain a more robust and more stringent constraint on $L/M$ from real data. We employ the X-ray reflection model {\sc relxill\_nk}~\cite{apj,prl}, which is currently the only available reflection model for testing the Kerr metric with Xspec~\cite{xspec}\footnote{http://heasarc.gsfc.nasa.gov/docs/xanadu/xspec/index.html}. We analyze the 2014 observations of \textsl{NuSTAR} and \textsl{Swift} of the supermassive black holes in 1H0707--495, and we fit the data to constrain the parameter $L/M$. Currently there are data of about twenty supermassive black holes and of about ten stellar-mass black holes that could be used to constrain $L/M$ with this technique. Here we consider 1H0707--495 because it is a source particularly suitable for our test. Its reflection component is very strong and the inner edge of its accretion disk is very close to the black hole: these are two ingredients that should help to get a stronger constraint on $L/M$. Our result is that $L/M < 0.45$ (90\% confidence level).

The present paper is organized as follows. In Section~\ref{s-x}, we briefly review the technique employed in our analysis, which is commonly called X-ray reflection spectroscopy. In Section~\ref{s-data}, we introduce the observations of \textsl{NuSTAR} and \textsl{Swift} that will be used for our study, and we describe how they have been reduced. In Section~\ref{s-ana}, we present the analysis of these observations, we show the results, and we discuss their physical implications. Our conclusions are summarized in Section~\ref{s-con}.


\section{X-ray reflection spectroscopy \label{s-x}}

X-ray reflection spectroscopy refers to the study of the reflection spectrum of the accretion disk around a black hole. In the disk-corona model~\cite{c1,c2}, a black hole accretes from a geometrically thin and optically thick disk. The disk is assumed to be in local thermal equilibrium, so any point of the disk emits as a blackbody at a certain temperature. The temperature profile of the accretion disk can be calculated from the Novikov-Thorne model~\cite{nt1,nt2}, which is the standard paradigm to describe geometrically thin and optically thick accretion disks around black holes. The temperature of the inner part of the accretion disk is generally in the soft X-ray band for stellar-mass black holes and in the optical/UV band for supermassive black holes. The ``corona'' is a hotter, typically $\sim 100$~keV, usually optically thin medium close to the black hole, but its exact geometry is currently not well understood. For instance, it may be the base of a jet, and in such a case it would be a compact source just above the black hole (lamppost corona). Another possibility is that the corona is an atmosphere covering the accretion disk (sandwich corona). Thermal photons from the accretion disk can inverse Compton scatter off free electrons in the corona. This process generates a power-law spectrum with a cut-off energy that depends on the corona temperature. A fraction of the X-ray photons of this power-law component illuminate the disk, producing a reflection component with some emission lines.

The strongest feature of the reflection spectrum is usually the iron K$\alpha$ line complex. In the rest-frame of the disk particles, the iron K$\alpha$ line is a very narrow feature at 6.4~keV in the case of neutral or weakly ionized iron, and it shifts up to 6.97~keV in the case of H-like iron ions. In the spectrum of an astrophysical black hole, this line can instead be very broad and skewed, as a result of relativistic effects occurring in the strong gravity region (gravitational redshift, Doppler boosting, light bending); for a review, see, for instance, Ref.~\cite{book}. X-ray reflection spectroscopy is sometimes called the iron line method because the iron K$\alpha$ line is usually the strongest feature in the disk's reflection spectrum, but in general a full reflection spectrum is necessary for fitting to data, not just a broadened iron line.

In the past 10-15~years, X-ray reflection spectroscopy has been developed to measure black hole spins under the assumption of the Kerr background~\cite{k1,k2,k3}. More recently, some authors have explored the possibility of using this technique to test Einstein's gravity in the strong field regime~\cite{nk1,nk2,nk3,nk4,nk5,nk6,nk7,nk8,nk9}. In the presence of the correct astrophysical model and of high quality data, X-ray reflection spectroscopy promises to be a powerful tool for probing the strong gravity region around both stellar-mass and supermassive black holes.

Currently, the most advanced X-ray reflection model for the Kerr metric is the {\sc relxill} package~\cite{re1,re2,re3,re4}. {\sc relxill\_nk} is its extension to non-Kerr spacetimes~\cite{apj,prl}. In Ref.~\cite{apj}, the model employed the Johannsen metric~\cite{j-m} and, in Ref.~\cite{prl}, we applied the model with the Johannsen metric to some data of the supermassive black hole in 1H0707--495 to constrain the deformation parameter $\alpha_{13}$ of the Johannsen spacetime. For the present paper, we have constructed a new version of {\sc relxill\_nk}, in which the geometry of the spacetime is described by the singularity-free black hole solution shown in Eq.~(\ref{eq-ds}). The construction of the whole model is similar to the work described in Ref.~\cite{apj}. The final result is a reflection model for testing Einstein's conformal gravity. In the next sections, we will employ our new model to analyze the \textsl{NuSTAR} and \textsl{Swift} data from 2014 of 1H0707--495 and constrain the dimensionless parameter $L/M$.


\begin{table}[t]
\centering
\begin{tabular}{ccccccc}
\hline
Mission & \hspace{0.2cm} & Obs.~ID & \hspace{0.2cm} & Year & \hspace{0.2cm} & Exposure time (ks) \\
\hline
 \textsl{NuSTAR} && 60001102002 && 2014 && 144 \\
 && 60001102004 && 2014 && 49 \\
 && 60001102006 && 2014 && 47 \\
\hline
 \textsl{Swift} && 00080720001 && 2014 && 20 \\
 && 00080720003 && 2014 && 17 \\
 && 00080720004 && 2014 && 17 \\
\hline
\end{tabular}
\vspace{0.3cm}
\caption{\textsl{NuSTAR} and \textsl{Swift} observations of 1H0707--495. In our analysis, we have not included the second \textsl{Swift} observation because it was taken during an anomaly period of this mission. \label{t-obs}}
\end{table}

\section{Observations and data reduction \label{s-data}}

1H0707--495 is classified as a Narrow Line Seyfert~1 galaxy. The X-ray spectrum of its nucleus has significant edge features, which are commonly interpreted as an extremely strong reflection component of the central supermassive black hole. With this interpretation and under the assumption of the Kerr metric, several authors have found the inner edge of the accretion disk to be very close to the black hole (which increases the relativistic effects), a moderate inclination angle, and an extremely high iron abundance~\cite{0707,0707bbb,fabian11,0707td12,kara15}. In the present work, we will adopt this interpretation that the spectrum of 1H0707--495 is reflection dominated and we will constrain the dimensionless parameter $L/M$. See, however, Ref.~\cite{hagino}, in which the authors suggest that the spectrum of 1H0707--495 is instead dominated by a powerful wind.

In archive, there are three separated observations of \textsl{NuSTAR} in 2014 with simultaneous snapshots of \textsl{Swift}/XRT. These observations are shown in Tab.~\ref{t-obs}. In our analysis we do not include the second \textsl{Swift} observation because it was taken during an anomaly period of this mission.

The \textsl{NuSTAR} data from both the FPMA and FPMB instruments were processed using nupipeline v0.4.5 with the standard filtering criteria and the NuSTAR CALDB version 20170120. For the spectra and light-curves extraction, we used the task {\it nuproduct}. We chose a circular source region with a radius of 40~arcsec and background region with a radius of 85~arcsec on the same chip. We did not find any pile-up effect in these \textsl{NuSTAR} observations. All spectra were binned to a minimum of 1 count before analysis. The \textsl{Swift}/XRT spectra were extracted following the standard criteria with a source region with a radius of 20~arcsec using the {\it xselect} tool. The data were binned to a minimum of 1 count in order to do a simultaneous fitting with the \textsl{NuSTAR} data. Because of the low photon count per bin, we used the Cash-statistics.


\section{Spectral analysis \label{s-ana}}

The \textsl{NuSTAR} and \textsl{Swift} data of 2014 of 1H0707--495 were analyzed for the first time in Ref.~\cite{kara15}. The spectrum can be described by a simple reflection model. We thus fit the data with the model {\sc tbabs*relxill\_nk}, where {\sc tbabs} takes into account the galactic absorption and {\sc relxill\_nk} describes the power-law component from the corona and the reflection spectrum from the disk. The data do not require any distant reflector or absorber. In our analysis, we impose that the values of the model parameters are the same for the three observations, with the exception of the photon index $\Gamma$, as done in~\cite{kara15}. The emissivity profile of the accretion disk is modeled with a simple power-law, i.e. $\propto 1/r^q$ where $q$ is the emissivity index and is left free in the fit. Because of the low photon count, we use the Cash-statistics.

\begin{figure}[t]
\begin{center}
\includegraphics[type=pdf,ext=.pdf,read=.pdf,width=9.0cm]{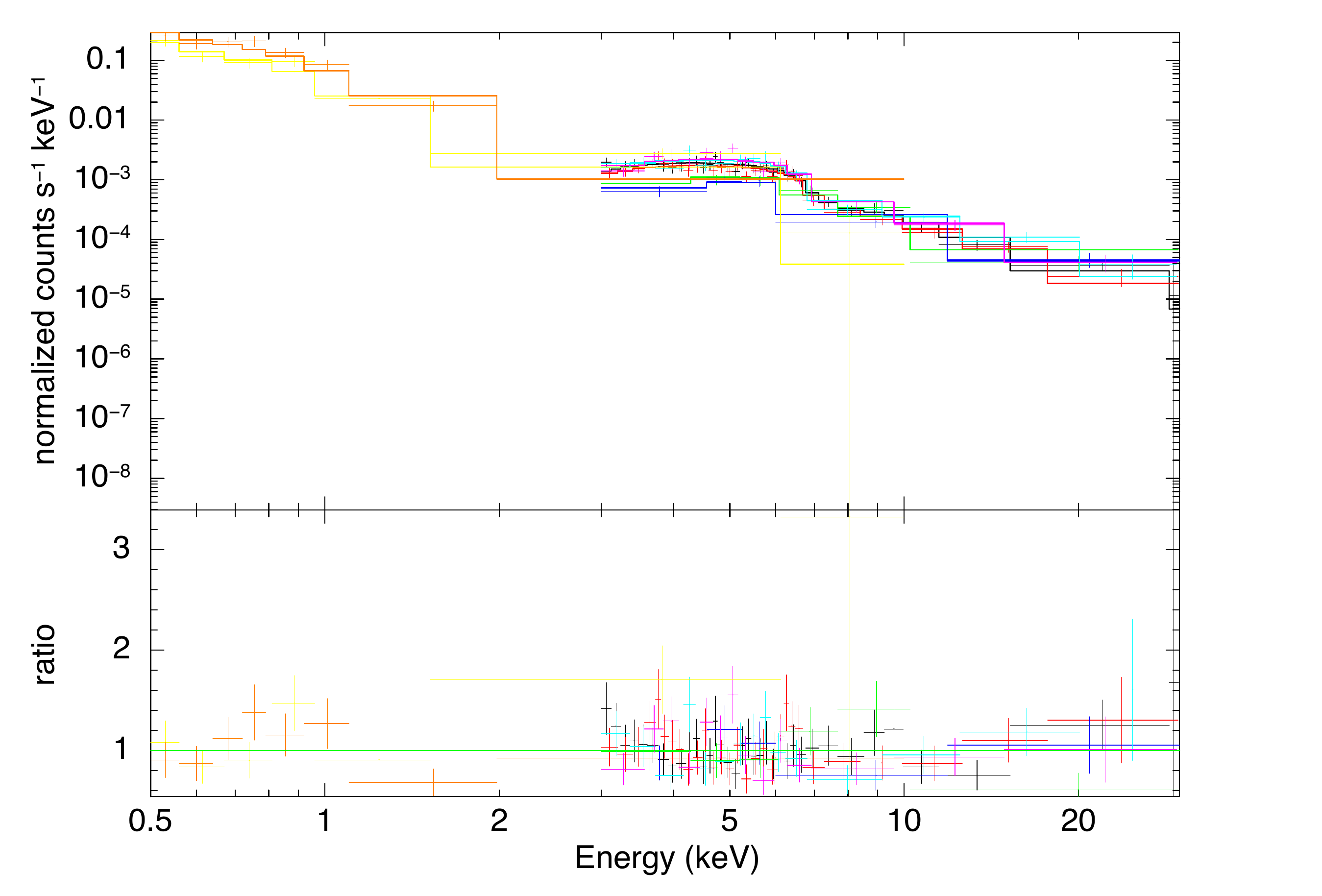}
\end{center}
\vspace{-0.5cm}
\caption{Data (top panel) and data-to-model ratio (bottom panel) of the \textsl{NuSTAR} and \textsl{Swift} observations of 1H0707--495 in 2014. The \textsl{NuSTAR} data of FPMA are in magenta, black, and green. The \textsl{NuSTAR} data of FPMB are in pale blue, red, and blue. The \textsl{Swift} data are in orange and yellow. The data have been rebinned for plotting purposes only. \label{f-ratio}}
\end{figure}

\begin{figure}[t]
\begin{center}
\hspace{-1.0cm}
\includegraphics[type=pdf,ext=.pdf,read=.pdf,width=9.0cm]{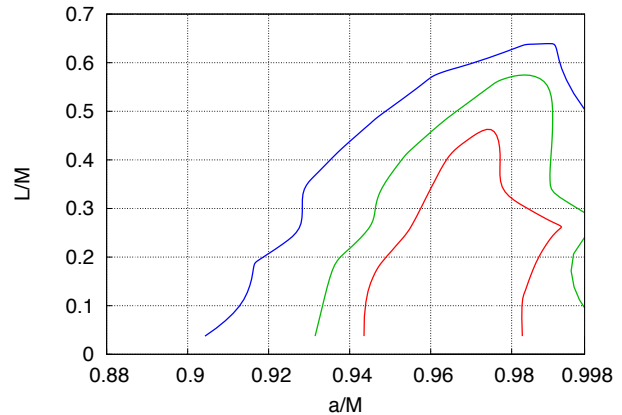}
\end{center}
\vspace{-1.1cm}
\caption{Constraints on the specific spin $a$ and the parameter $L$ from the \textsl{NuSTAR} and \textsl{Swift} data of the supermassive black hole in 1H0707--495. The red, green, and blue lines indicate, respectively, the 68\%, 90\%, and 99\% confidence level contours for two relevant parameters. The contours do extend to $L/M=0$, although they look to be open-ended in the plot. See the text for more details. \label{f-plot}}
\end{figure}

Table~\ref{t-fit} shows the best-fit values. The three $\Gamma_i$'s correspond to the photon index of the three observations. The iron abundance $A_{\rm Fe}$ (expressed in units of the Solar iron abundance) is particularly high, but this is a well known result for this source when we interpret its spectrum as reflection dominated~\cite{0707,0707bbb,fabian11,0707td12,kara15}. The spin parameter is high and we find the constraint
\be
L/M < 0.45 \quad \text{(90\% C.L. for one relevant parameter)} \, . \nonumber
\ee
Figure~\ref{f-ratio} shows the data (top panel) as well as the data to model ratio (bottom panel). Figure~\ref{f-plot} shows the constraints on the specific spin $a$ and on the parameter $L$. The red, green, and blue lines indicate, respectively, the 68\%, 90\%, and 99\% confidence level contours for two relevant parameters.

\begin{table}[t]
\centering
\begin{tabular}{lcc}
\hline\hline
Model parameters & \hspace{0.5cm} & Best-fit values \\
\hline
$a/M$ && $> 0.94$ \\
$L/M$ && $< 0.45$ \\
$i$ [deg] && $41^{+2}_{-2}$ \\
$q$ && $4.2^{+0.5}_{-0.2}$ \\
$\Gamma_1$ && $3.28^{+0.02}_{-0.05}$ \\
$\Gamma_2$ && $2.58^{+0.10}_{-0.06}$ \\
$\Gamma_3$ && $3.12^{+0.03}_{-0.05}$ \\
$\log\xi$ && $2.08^{+0.22}_{-0.05}$ \\
$A_{\rm Fe}$ && $>9.1$ \\
\hline
C-stat/dof && 1940.58/3246 \\
\hline\hline
\end{tabular}
\vspace{0.3cm}
\caption{Summary of the best-fit values. The reported uncertainties correspond to the 90\% confidence level for one relevant parameter. Note we have employed the Cash-statistics because of the low photon count. See the text for more details. \label{t-fit}}
\end{table}

\section{Concluding remarks \label{s-con}}

Einstein's gravity has been extensively tested in weak gravitational fields, and current data agree with its theoretical predictions. However, there are many alternative theories of gravity that have the same predictions as Einstein's gravity for weak fields and present deviations only in the strong gravity regime. Astrophysical black holes are the ideal laboratory for testing Einstein's gravity in the strong field regime.

In the present paper, we have considered the black hole solutions of a large family of conformal theories of gravity~\cite{p1}. In these solutions, the spacetime metric is characterized by the black hole mass $M$, the black hole spin angular momentum $J$, and a parameter $L$ with the dimensions of length. The theory does not specify the value of $L$, but it is natural to expect that $L$ is either of the order of the Planck length or of the order of the mass of the black hole $M$. For $L = 0$, the solution reduces to the Kerr metric of Einstein's gravity. Here we have focused on the second scenario with $L \propto M$ and have obtained a constraint on $L/M$ from the analysis of the reflection spectrum of the accretion disk around the supermassive black hole in 1H0707--495.

We have applied the recent X-ray reflection model {\sc relxill\_nk}~\cite{apj} to the 2014 observations of \textsl{NuSTAR} and \textsl{Swift} of 1H0707--495. Assuming that the spectrum is reflection dominated, we have obtained the constraint $L/M < 0.45$ (90\% confidence level for one relevant parameter). Our measurement is thus consistent with the Kerr metric, but the constraint is not so stringent as to rule out the possibility of a non-vanishing $L$ of the order of $M$.


\begin{acknowledgments}
This work was supported by the National Natural Science Foundation of China (NSFC), Grant No.~U1531117, and Fudan University, Grant No.~IDH1512060. A.A. also acknowledges the support from the Shanghai Government Scholarship (SGS). C.B. also acknowledges support from the Alexander von Humboldt Foundation. S.N. also acknowledges support from the Excellence Initiative at Eberhard-Karls Universit\"at T\"ubingen.
\end{acknowledgments}


\end{document}